# Nonlinear Schrödinger equation and classical-field description of the Lamb-Retherford experiment

Sergey A. Rashkovskiy

*Institute for Problems in Mechanics of the Russian Academy of Sciences, Vernadskogo Ave., 101/1, Moscow, 119526, Russia*

*Tomsk State University, 36 Lenina Avenue, Tomsk, 634050, Russia*

*E-mail: rash@ipmnet.ru, Tel. +7 906 0318854*

**Abstract**
I show that Lamb-Retherford experiment can be fully described within the framework of classical field theory without using concepts such as the discrete states of the atom and jump-like electron transitions between them. The rate of stimulated decay of the metastable state of a hydrogen atom in an external periodic electric field is determined. The dependence of this rate on the frequency and amplitude of the external electric field, as well as on the parameters of the atom, has been obtained. It is shown that the maximum value of the stimulated decay rate of the metastable state of a hydrogen atom is achieved at an external electric field frequency equal to the frequency shift corresponding to either the fine structure of the hydrogen atom or the Lamb shift.

## 1. Introduction

The experiment of Lamb and Retherford [1] is considered one of the most elegant experiments in the history of physics.

Using the microwave stimulation of excited hydrogen atoms, Lamb and Retherford were able to observe transitions between the levels $2s_{1/2} - 2p_{1/2}$ and $2s_{1/2} - 2p_{3/2}$ which are degenerate in accordance with the Schrödinger equation.

These experiments made it possible to discover a phenomenon such as the Lamb shift, and quantitatively investigate the fine structure of the spectral lines of the hydrogen atom.

In particular, the frequency separation for the fine structure $2s_{1/2} - 2p_{3/2}$, was found to be 10949 MHz. For the Lamb shift, Lamb and Retherford found in 1947 that the $2s_{1/2}$ level lies above the $2p_{1/2}$ level by an amount of about 1000 MHz [1]. Further experiments carried out by Lamb and his collaborators [2-6] gave the very precise value (1057.77±0.10) MHz for this frequency difference. This value is about one-tenth of the fine structure splitting of the $n = 2$ term.

The experimental method of Lamb and Retherford is based on the fact that the $2s_{1/2}$ level is metastable. Indeed, the spontaneous electric dipole transition from the state $2s_{1/2}$ to the ground state $1s_{1/2}$ is forbidden by the selection rule $\Delta l = \pm 1$. The most probable decay mechanism of the $2s_{1/2}$ state is a quadrupole emission (so-called two-photon emission), with lifetime of 1/7 s. Thus, in the absence of perturbations, the lifetime of the $2s_{1/2}$ state is very long compared to that of the $2p$ states, which is about $1.6 \times 10^{-9}$ s. In the apparatus of Lamb and Retherford, a beam of atomic hydrogen containing atoms excited to the metastable $2s_{1/2}$ state was produced by first dissociating molecular hydrogen in a tungsten oven (at a temperature of 2500 K where the dissociation is about 64 per cent complete), selecting a jet of atoms by means of slit, and bombarding this jet with a beam of electrons having a kinetic energy somewhat larger than 10.2 eV, which is the threshold energy for excitation of the $n = 2$ levels of atomic hydrogen [7]. As a result, a small fraction of hydrogen atoms (about one in $10^8$) was excited to the $2s_{1/2}$, $2p_{1/2}$ and $2p_{3/2}$ states. The average velocity of the atomic beam was about $8 \times 10^3$ m/s. Because of their long lifetime, the atoms excited to the metastable $2s_{1/2}$ state could easy reach a detector placed at a distance of about 10 cm from the region where they were produced. On the other hand the atoms which were excited in the $2p_{1/2}$ or $2p_{3/2}$ states quickly decayed to the ground state $1s_{1/2}$ in $1.6 \times 10^{-9}$ s, moving only about $1.3 \times 10^{-3}$ cm in that time, so they could not reach the detector. This detector was a tungsten foil, from which the atoms in the metastable state $2s_{1/2}$ could eject electrons by giving up their excitation energy. Atoms in the ground state cannot produce electronic current; because of this the measured electronic current was proportional to the number of metastable atoms reaching the detector. The beam containing the metastable $2s_{1/2}$ atoms was passed through an "interaction region" in which a microwave field of proper frequency was applied. Under action of microwave field, the metastable atoms underwent induced transitions to the $2p_{1/2}$ or $2p_{3/2}$ states, and quickly decayed to the ground state $1s_{1/2}$ in which they could not be detected. As a result, it was observed a reduction of the number of metastable ($2s_{1/2}$) atoms registered by the detector at the (resonant) microwave frequencies corresponding to the frequencies of the $2s_{1/2} - 2p_{1/2}$ and $2s_{1/2} - 2p_{3/2}$ transitions. In the "interaction region" the atomic beam also passed in a variable magnetic field. This field was used by Lamb and Retherford not only to separate the Zeeman components of the $2s_{1/2}$, $2p_{1/2}$ and $2p_{3/2}$ levels, but also to reduce the probability of fortuitous disintegration of the $2s_{1/2}$ state due to Stark effect mixing of the $2s_{1/2}$ and the $2p$ levels caused by perturbing fields. Moreover, the use of a variable external magnetic field has made it possible to avoid the difficulty of producing a microwave field with a variable frequency but a constant microwave power. Instead, Lamb and Retherford could operate at a fixed frequency of the microwave field and obtained the passage through the resonance by varying the magnetic field. The resonance frequency for zero magnetic field was found by extrapolation.

The explanation of the Lamb-Retherford experiment is usually based on the concepts traditional for quantum mechanics about the existence of discrete energy levels of an atom and jump-like transitions of electrons between them, as well as on the existence of forbidden transitions.

At the same time, it was shown in [8-11] that all the basic properties of the hydrogen atom can be easily explained within the framework of classical field theory if the electron is considered not as a particle which has the properties unusual from the classical point of view but as an actual classical field described by the wave function $\Psi$. From this point of view, the Dirac equation and its reduced forms (the Pauli, Schrödinger and Klein-Gordon equations) are field equations for the electron field $\Psi$, just as the Maxwell equations are field equations for the classical electromagnetic field.

This point of view fully agrees with the classical concepts of fields and particles and does not require the postulating of any special "quantum", nonclassical properties of the electromagnetic field and the "electron" [8-14].

In particular, as shown in [9], from this point of view, the hydrogen atom can be regarded as a classical open volume resonator in which the electron wave is held due to "total internal reflection", while the Coulomb field of the proton plays the role of the dielectric medium. The electron wave in a hydrogen atom is a completely classical continuous (not quantized) field, analogous to a classical electromagnetic field, and does not reduce to any particles or quanta. Unusual in comparison with the classical electromagnetic field is that the electron wave has an electric charge, its own intrinsic angular momentum and the associated intrinsic magnetic moment, continuously distributed in space [8]. The last two properties of an electron wave form what is called a spin in quantum mechanics [8]. Similar to any classical resonator, the hydrogen atom has eigenmodes and the corresponding eigenfrequencies $\omega_n$ of the electron wave, which are, respectively, eigenfunctions and eigenvalues of the linear Schrödinger (Dirac, Pauli, Klein-Gordon) equation. Using such representations, it was shown in [9] that the spontaneous emission of a hydrogen atom has a simple classical physical meaning, and is associated with the oscillations of the distributed electric charge of the electron wave. If the electron wave is in only one eigenmode (the pure state of the atom), it has a stationary electric charge distribution and, according to classical electrodynamics, does not emit electromagnetic waves. In quantum mechanics this is called the stationary state of the atom. If several eigenmodes are excited simultaneously in a hydrogen atom (mixed state of the atom), oscillations of the distributed electric charge of the electron wave arise, which according to classical electrodynamics leads to the emission of electromagnetic waves, and the frequencies of the emitted electromagnetic waves are equal to the difference between the eigenfrequencies of the excited modes of the electron wave in the hydrogen atom.

As shown in [9], such a point of view allows giving a noncontradictory classical description of the spontaneous emission of an atom and to obtain in a natural way many of the dependencies traditionally derived in the framework of quantum electrodynamics. In particular, this theory provides a natural explanation for the selection rule and the existence of so-called "forbidden" transitions; these conclusions are a natural consequence of classical electrodynamics.

In the traditional interpretation of quantum mechanics, spontaneous emission is associated with jump-like quantum transitions of an electron between discrete levels $n$ and $k$.

According to theory [8-11], no jump-like transitions occur; on the contrary, there is a continuous flow of electric charge from the excited mode $n$ into the excited mode $k$ (hereinafter it is assumed that $\omega_n > \omega_k$), accompanied by continuous spontaneous emission of electromagnetic waves. Thus, an electron wave can be simultaneously in two or more excited modes, but in this case it will spontaneously and continuously "flow" into mode with a lower eigenfrequency if this transition is not forbidden by the selection rule.

In those cases when the electric dipole moment of an electron wave, which is in a mixed state, remains constant, the electric dipole spontaneous emission of the atom does not occur; this is a trivial consequence of classical electrodynamics. In quantum mechanics, this is traditionally interpreted as "forbidden transitions".

Note that in the absence of electric dipole spontaneous emission, multipole (for example, quadrupole) and magnetic dipole (associated with both "orbital motion" and spin) emissions can occur, which, however, have a significantly lower intensity in comparison with electric dipole emission, and, therefore, lead to a slower loss of energy.

In the traditional interpretation of quantum mechanics, there is the problem of explaining the reason for the spontaneous transition: what makes an atom which is in a stationary excited state go into another stationary state with a lower energy level? The linear Schrödinger (Dirac, Pauli, Klein-Gordon) equation does not give an answer to this question, because it cannot describe the spontaneous transition. For this reason, an additional hypothesis that the spontaneous transition is associated with the impact of zero-point fluctuations on the electron, which "swing" the electron (so-called zitterbewegung) and cause it to "jump" into a stationary state with lower energy is introduced in quantum mechanics.

Within the framework of the theory [8-11], there are no problems with the explanation of the spontaneous transition, which is a trivial consequence of classical electrodynamics, as a result of the inverse action of the self-radiation field on atom. The taking into account the self-radiation field leads to the appearance of nonlinear terms in the Schrodinger equation, which describe the change in the structure of the hydrogen atom in the process of spontaneous emission [9]. In this paper we show that Lamb-Retherford experiments can also be fully described in the framework of classical field theory [8-11] without using the representations about discrete states of the atom and jump-like electron transitions between them.

From the point of view of the theory [8-11], the Lamb-Retherford experiment has a simple classical explanation. Suppose that three eigenmodes are simultaneously excited in the hydrogen atom, for example, $1s_{1/2}$, $2s_{1/2}$ and $2p_{1/2}$, which corresponds to the experiment for determining the Lamb shift. The mixed state of an atom with two excited modes $2p_{1/2}$ and $2s_{1/2}$ is nonstationary, but the lifetime of this state is very long (of the order of several years) because of the very small difference in frequencies of the modes $2p_{1/2}$ and $2s_{1/2}$ (the rate of relaxation of mixed state is proportional to the third power of the frequency difference). According to classical electrodynamics, there is no electric dipole emission if an atom is in the mixed state with two excited modes $2s_{1/2}$ and $1s_{1/2}$, since the electric dipole moment in this state is constant [9-11].

The mixed state of an atom with two excited modes $2p_{1/2}$ and $2s_{1/2}$ is nonstationary, but the lifetime of this state is very long (of the order of several years) because of the very small difference in frequencies of the modes $2p_{1/2}$ and $2s_{1/2}$ (the rate of relaxation of mixed state is proportional to the third power of the frequency difference). According to classical electrodynamics, there is no electric dipole emission if an atom is in the mixed state with two excited modes $2s_{1/2}$ and $1s_{1/2}$, since the electric dipole moment in this state is constant [9-11]. If the atom for some reason is in a mixed state with two excited modes $2s_{1/2}$ and $1s_{1/2}$, it will slowly decay to the ground state due to quadrupole radiation during ~ 1/7 s. The mixed state with two excited modes $2p_{1/2}$ and $1s_{1/2}$ is nonstationary with a very short lifetime ~$1.6 \times 10^{-9}$ s. Because of this, if for some reason the $2p_{1/2}$ mode will turn out to be even weakly excited (in the Lamb-Retherford experiments, the ground mode $1s_{1/2}$ is always excited), then a rapid spontaneous flow of the electron wave (and its electric charge) from this mode into the ground mode $1s_{1/2}$ will begin. As a result, the $2p_{1/2}$ mode will "empty" in a time of the order of $1.6 \times 10^{-9}$ s, i.e. almost instantaneously in comparison with the lifetime of the mixed state $(2s_{1/2}; 1s_{1/2})$.

At the same time, if an atom that is in a mixed state with two excited modes $2s_{1/2}$ and $1s_{1/2}$ is placed in an alternating electric field oscillating with a frequency equal to the difference of the eigenfrequencies of the $2p_{1/2}$ and $2s_{1/2}$, the induced flow of the electron wave from the $2s_{1/2}$ mode into the $2p_{1/2}$ mode will begin, from which it quickly and spontaneously will flow into the ground mode $1s_{1/2}$. A similar process will occur if such an atom is placed in an alternating electric field that oscillates with a frequency equal to the difference between the eigenfrequencies of the $2p_{3/2}$ and $2s_{1/2}$ modes. In this case, the induced flow of the electron wave from the $2s_{1/2}$ mode into the $2p_{3/2}$ mode will begin, from which it quickly spontaneously will flow into the ground mode $1s_{1/2}$. This was realized in the Lamb-Retherford experiments. The theory [8-12] allows quantitatively describing the Lamb-Retherford experiments.

## 2. Description of the Lamb shift by the Dirac equation

The hydrogen atom is described by the Dirac equation
$$i\hbar \frac{\partial \Psi}{\partial t} = \widehat{H}_D \Psi \qquad (1)$$
where $\Psi$ is the bispinor; the operator $\widehat{H}_D$ includes both the electrostatic field of the nucleus and the external electromagnetic field in which the atom can be located.

The Dirac equation (1) in the absence of an external electromagnetic field removes the degeneracy associated with the fine structure of the spectral lines, but does not remove the degeneracy associated with the Lamb shift [15]. For example, according to equation (1), the frequencies of eigenmodes $2s_{1/2}$ and $2p_{1/2}$ coincide, although in reality they differ by ~ 1057.85 MHz (Lamb shift). As is known, the Lamb shift is calculated in the framework of quantum electrodynamics as a radiation correction, but formally it can be described by the Dirac equation if we add to it the term $\widehat{V}\Psi$ describing some additional effect that removes the last degeneracy of the eigenmodes of the hydrogen atom:
$$i\hbar \frac{\partial \Psi}{\partial t} = \widehat{H}_D \Psi + \widehat{V} \Psi \qquad (2)$$
In order for the modified Dirac equation (2) to have stationary solutions $\Psi(t,\mathbf{r}) = u_n(\mathbf{r})\exp(-i\omega_n t)$, the operator $\widehat{V}$ must satisfy the condition $\widehat{V}[u_n \exp(-i\omega_n t)] = \exp(-i\omega_n t)\widehat{V}_n u_n$, where $\widehat{V}_n$ is the operator acting on the spatial coordinates and on the spinor indices.

Thus, with the appropriate choice of the operator $\widehat{V}_n$, the stationary modified Dirac equation
$$\hbar \omega_n u_n = \widehat{H}_D^0 u_n + \widehat{V}_n u_n \qquad (3)$$
will have correct eigenvalues $\omega_n$ (including the fine structure and Lamb shift), where $\widehat{H}_D^0$ is the Dirac's Hamiltonian of the hydrogen atom in the absence of an external field, while the term $\widehat{V}\Psi$ (and, respectively, $\widehat{V}_n u_n$) can be considered as a small perturbation.

Here we will not discuss the nature of the additional impact on the atom, which is described by the operator $\widehat{V}$ (this is planned in the next papers in the framework of the theory [8-11]); we assume that it is known and gives the correct values of the eigenfrequencies for all eigenmodes of the hydrogen atom, including the Lamb shift, in particular, it removes the degeneracy of the $2s_{1/2}$ and $2p_{1/2}$ modes.

Then in an external field the hydrogen atom will be completely described by the equation (2) with the corresponding operator $\hat{V}$. Because of the smallness of the action described by the operator $\hat{V}$ (which can be regarded as a small perturbation), one can neglect the influence of the external field (which can also be regarded as a small perturbation) on the operator $\hat{V}$. In other words, the operator $\hat{V}$ can be considered the same, both in the absence of an external field, and in its presence.

As we show below, it is sufficient to use the long-wave [8] (or, as it is customary, nonrelativistic) approximation of equation (2) to describe the Lamb-Retherford experiment.

We write equation (2) in the long-wave approximation, preserving such effects as the fine structure of the spectral lines and the Lamb shift. In this case, we follow the style of presentation [15].

The wave function (bispinor) in the long-wave approximation can be represented in the form

$$\Psi = \begin{pmatrix} \psi \exp(-i\omega_e t) \\ \chi \exp(-i\omega_e t) \end{pmatrix} \qquad (4)$$

where $\omega_e = \frac{m_e c^2}{\hbar}$; $m_e$ is the electron mass; $\psi$ and $\chi$ are the spinors.

In this case, the electric charge density of an electron wave in a hydrogen atom can be written in the form

$$\rho = -e\Psi^*\Psi = -e(\psi^*\psi + \chi^*\chi) \qquad (5)$$

We consider the standard representation of the wave function, which is usual in such cases, for which, in the long-wave approximation, $\chi \ll \psi$ [15].

The transition to the long-wave approximation from the mathematical point of view means, in fact, the expansion of the wave function $\Psi$ in a series in the small parameter $\alpha = \frac{e^2}{\hbar c} \ll 1$, which is a fine-structure constant. In this case, $\psi = \psi_0 + \alpha\psi_1 + \alpha^2\psi_2 + \cdots$; $\chi = \alpha\chi_1 + \alpha^2\chi_2 + \cdots$.

In the first (Schrödinger) long-wave approximation, only terms of not higher than first order in $\alpha$ remain. In this case

$$\chi = \frac{1}{2m_e c}\boldsymbol{\sigma}\left(\frac{\hbar}{i}\nabla + \frac{e}{c}\mathbf{A}\right)\psi, \qquad (6)$$

the electric charge density of an electron wave in a hydrogen atom is determined by the Schrödinger expression

$$\rho = -e|\psi|^2, \qquad (7)$$

while equation (2) reduces to the Pauli equation

$$i\hbar\frac{\partial\psi}{\partial t} = \hat{H}_{SP}\psi \qquad (8)$$

with respect to the spinor $\psi$, where

$$\hat{H}_{SP} = \frac{1}{2m_e}\left(\frac{\hbar}{i}\nabla + \frac{e}{c}\mathbf{A}\right)^2 - \frac{e^2}{r} + \frac{e\hbar}{2m_e c}\boldsymbol{\sigma}\mathbf{H} \tag{9}$$

is the Schrödinger-Pauli Hamiltonian ($e > 0$).

In the next (post-Schrödinger) approximation, the terms in the expansion of the wave function, of the order of $\alpha^2$, are taken into account. In this case, the density of the electric charge of the electron wave in the hydrogen atom (5), with allowance for (6), can be written in the form [15]

$$\rho = -e\left(|\psi|^2 + \frac{\hbar^2}{4m_e^2 c^2}|\boldsymbol{\sigma}\nabla\psi|^2\right) \tag{10}$$

This expression differs from Schrödinger's one (7).

At the same time, instead of the function $\psi$, another two-component function can be introduced [15]

$$\psi_S = \left(1 - \frac{\hbar^2}{8m_e^2 c^2}\Delta\right)\psi, \quad \psi = \left(1 + \frac{\hbar^2}{8m_e^2 c^2}\Delta\right)\psi_S \tag{11}$$

such that in the post-Schrödinger approximation under consideration the integral

$$\int |\psi_S|^2 dV = \int\left(|\psi|^2 + \frac{\hbar^2}{4m_e^2 c^2}|\boldsymbol{\sigma}\nabla\psi|^2\right) dV = 1 \tag{12}$$

persists in time. Then, for the electric charge density of an electron wave in a hydrogen atom, we can write

$$\rho \approx -e|\psi_S|^2 \tag{13}$$

Function $\psi_S$ satisfies the equation [15]

$$i\hbar\frac{\partial \psi_S}{\partial t} = \left(\hat{H}_{SP} + \hat{U}\right)\psi_S \tag{14}$$

where

$$\hat{U} = -\frac{\hbar^4}{8m_e^3 c^2}\Delta^2 - i\frac{e\hbar^2}{4m_e^2 c^2}\boldsymbol{\sigma}[\mathbf{E}\nabla] + \frac{e\hbar^2}{8m_e^2 c^2}\text{div}\mathbf{E} \tag{15}$$

$\mathbf{E}$ is the electric field strength (both the electrostatic field of the proton and the external field). The operator (15) describes the fine structure of the spectral lines of the hydrogen atom [15]. It is a small perturbation to the Schrödinger-Pauli Hamiltonian (9). If the external field in which the hydrogen atom is located is also a small perturbation (as is the case in the Lamb-Retherford experiments), then in the operator (15), it can be ignored, because it will give effects of a higher order which are of no interest to us in this work. For this reason, only the electrostatic field of the proton should be considered in operator (15), and it can be considered the same both in the absence of an external field and in its presence, similarly to the operator $\hat{V}$. Let us consider the result of the action of the operator $\hat{V}$ in the post-Schrödinger long-wave approximation. To do this, we write it in the form

$$\hat{V} = \begin{pmatrix} \hat{V}_{11} & \hat{V}_{12} \\ \hat{V}_{21} & \hat{V}_{22} \end{pmatrix} \tag{16}$$

where $\hat{V}_{ij}$ are the corresponding operators.

In the approximation under consideration, equation (2) with allowance for equations (6) and (11) can be written in the form

$$i\hbar \frac{\partial \psi_S}{\partial t} = (\hat{H}_{SP} + \hat{U} + \hat{W})\psi_S \tag{17}$$

where we introduced the operator

$$\hat{W} = \left[\hat{V}_{11} - i\frac{1}{2m_e c}\hat{V}_{12}(\boldsymbol{\sigma}\nabla)\right]\left(1 + \frac{\hbar^2}{8m_e^2 c^2}\Delta\right) \tag{18}$$

If the hydrogen atom is in an external electric field $\mathbf{E} = -\nabla\Phi$, $\mathbf{A} = 0$, where $\Phi$ is the scalar potential of the external field, then it is described by the equation (17), (18), which takes the form (hereinafter, the index "$S$" of the function $\psi_S$ will be omitted, this should not cause confusion, since we will use only the function $\psi_S$)

$$i\hbar \frac{\partial \psi}{\partial t} = \left(-\frac{1}{2m_e}\Delta - \frac{e^2}{r} - e\Phi + \hat{U} + \hat{W}\right)\psi \tag{19}$$

## 3. Eigenmodes of the hydrogen atom

In the absence of an external field ($\Phi = 0$), the solution of equation (19) can be written in the form

$$\psi(t, \mathbf{r}) = \sum_n c_n u_n(\mathbf{r}) \exp(-i\omega_n t) \tag{20}$$

where

$$u_n(\mathbf{r}) = \begin{pmatrix} u_{n,1} \\ u_{n,2} \end{pmatrix} = \sum_s B_{ns} w_s(\mathbf{r}) \tag{21}$$

are the two-component functions;

$$B_{ns} = \begin{pmatrix} a_{ns} \\ b_{ns} \end{pmatrix} \tag{22}$$

are the two-component parameters; $c_n$ are the complex parameters; $\omega_n$ are the frequency; functions $w_n(\mathbf{r})$ are eigenfunctions of the stationary Schrödinger equation

$$\hbar\omega_n^{(0)} w_n = \left(-\frac{1}{2m_e}\Delta - \frac{e^2}{r}\right)w_n \tag{23}$$

$\omega_n^{(0)}$ are the eigenvalues of the equation (23). For some $n$, the eigenvalues $\omega_n^{(0)}$ coincide (degenerate modes).

The eigenfunctions $w_n(\mathbf{r})$ form an orthonormal system:

$$\int w_n(\mathbf{r}) w_k^*(\mathbf{r}) d\mathbf{r} = \delta_{nk} \tag{24}$$

Two-component parameters $B_{ns}$ satisfy the condition

$$\sum_s B_{ns} B_{ks}^* = \sum_s (a_{ns} a_{ks}^* + b_{ns} b_{ks}^*) = \delta_{nk} \tag{25}$$

Taking into account (24) and (25), it is clear that the functions $u_n(\mathbf{r})$ also form an orthonormal system:

$$\int u_n(\mathbf{r})u_k^*(\mathbf{r})d\mathbf{r} = \int u_{n,1}(\mathbf{r})u_{k,1}^*(\mathbf{r})d\mathbf{r} + \int u_{n,2}(\mathbf{r})u_{k,2}^*(\mathbf{r})d\mathbf{r} = \delta_{nk} \quad (26)$$

Since the function (20) must satisfy the normalization condition (12), then, using (26), we obtain

$$\sum_n |c_n|^2 = 1 \quad (27)$$

Substituting equation (20) into equation (19) for $\Phi = 0$, we obtain

$$\sum_n c_n \hbar\omega_n u_n(\mathbf{r})\exp(-i\omega_n t) =$$

$$\left(-\frac{1}{2m_e}\Delta - \frac{e^2}{r}\right)\sum_n c_n u_n(\mathbf{r})\exp(-i\omega_n t) + \sum_n c_n \exp(-i\omega_n t)(\widehat{U} + \widehat{W})u_n(\mathbf{r}) \quad (28)$$

Here it is taken into account that

$$(\widehat{U} + \widehat{W})[c_n u_n(\mathbf{r})\exp(-i\omega_n t)] = c_k \exp(-i\omega_n t)(\widehat{U} + \widehat{W})u_n(\mathbf{r}) \quad (29)$$

Assuming that all the frequencies $\omega_n$ are different (there is no degeneracy, since the last degeneracy is removed by the action of $\widehat{V}\Psi$), we obtain

$$\hbar\omega_n u_n(\mathbf{r}) = \left(-\frac{1}{2m_e}\Delta - \frac{e^2}{r}\right)u_n(\mathbf{r}) + (\widehat{U} + \widehat{W})u_n(\mathbf{r}) \quad (30)$$

Taking into account equations (21) and (23), one obtains

$$\sum_s B_{ns}\hbar(\omega_n - \omega_s^{(0)})w_s(\mathbf{r}) = \sum_s B_{ns}(\widehat{U} + \widehat{W})w_s(\mathbf{r}) \quad (31)$$

Multiplying equation (31) by $w_k^*(\mathbf{r})$, and integrating over the whole space, taking into account equation (24), we obtain

$$\hbar(\omega_n - \omega_k^{(0)})B_{nk} = \sum_s Q_{ks}B_{ns} \quad (32)$$

where

$$Q_{ks} = \langle w_k|(\widehat{U} + \widehat{W})|w_s\rangle \quad (33)$$

or in expanded form,

$$Q_{ks} = \begin{pmatrix} \langle w_k|(\widehat{U} + \widehat{W})_{11}|w_s\rangle & \langle w_k|(\widehat{U} + \widehat{W})_{12}|w_s\rangle \\ \langle w_k|(\widehat{U} + \widehat{W})_{21}|w_s\rangle & \langle w_k|(\widehat{U} + \widehat{W})_{22}|w_s\rangle \end{pmatrix} \quad (34)$$

Then the system of equations (32) splits into two related systems of equations

$$\hbar(\omega_n - \omega_k^{(0)})a_{nk} = \sum_s[(Q_{ks})_{11}a_{ns} + (Q_{ks})_{12}b_{ns}] \quad (35)$$

$$\hbar(\omega_n - \omega_k^{(0)})b_{nk} = \sum_s[(Q_{ks})_{21}a_{ns} + (Q_{ks})_{22}b_{ns}] \quad (36)$$

This is a system of linear homogeneous equations, and it has non-trivial solutions only if its determinant $\Delta$ is equal to zero: $\Delta = 0$. From this condition we can find all the eigenfrequencies $\omega_n$, and then, at known frequencies $\omega_n$ and condition (25), we can find all the parameters $a_{ns}$ and $b_{ns}$ are found.

However, taking into account that the perturbations $Q_{ks}$ to the Schrödinger equation are small, the parameters $\omega_n$, $a_{ns}$ and $b_{ns}$ can be easily found explicitly using the perturbation theory. For this purpose we introduce

$$B_{nk} = B_{nk}^{(0)} + B_{nk}^{(1)} \tag{37}$$

where

$$B_{nk}^{(0)} = \begin{pmatrix} a_{nk}^{(0)} \\ b_{nk}^{(0)} \end{pmatrix}; \quad B_{nk}^{(1)} = \begin{pmatrix} a_{nk}^{(1)} \\ b_{nk}^{(1)} \end{pmatrix} \tag{38}$$

the parameters $B_{nk}^{(0)}$ correspond to the solution of the unperturbed Schrödinger equation (i.e. $Q_{ks} = 0$), while the parameters $B_{nk}^{(1)}$ are small additives related to the perturbations $Q_{ks}$. The condition (25) is satisfied both for the "perturbed" parameters $B_{nk}$ and for the unperturbed parameters $B_{nk}^{(0)}$:

$$\sum_s B_{ns}^{(0)} B_{ks}^{(0)*} = \delta_{nk} \tag{39}$$

Taking into account equations (25), (39), and (37) in the approximation linear in the perturbations, we obtain

$$\sum_s \left( B_{ns}^{(1)} B_{ks}^{(0)*} + B_{ns}^{(0)} B_{ks}^{(1)*} \right) = 0 \tag{40}$$

In the same approximation, equation (32) takes the form:

$$\hbar(\omega_n - \omega_k^{(0)})(B_{nk}^{(0)} + B_{nk}^{(1)}) = \sum_s Q_{ks} B_{ns}^{(0)} \tag{41}$$

It follows that for those $k$ for which $\omega_n^{(0)} \neq \omega_k^{(0)}$, we obtain

$$B_{nk}^{(0)} = 0 \tag{42}$$

and

$$\hbar(\omega_n^{(0)} - \omega_k^{(0)}) B_{nk}^{(1)} = \sum_s Q_{ks} B_{ns}^{(0)} \tag{43}$$

For those $k$ for which $\omega_n^{(0)} = \omega_k^{(0)}$ (degenerate modes), we obtain

$$\hbar(\omega_n - \omega_k^{(0)}) B_{nk}^{(0)} = \sum_s Q_{ks} B_{ns}^{(0)} \tag{44}$$

Note that the sums (43) and (44) contain only those $B_{ns}^{(0)}$, which correspond to degenerate modes for which $\omega_n^{(0)} = \omega_s^{(0)}$ and $B_{ns}^{(0)} \neq 0$.

Equation (44) can be written in the form

$$\sum_s \left( Q_{ks} - \hbar(\omega_n - \omega_k^{(0)}) \delta_{ks} \right) B_{ns}^{(0)} = 0 \tag{45}$$

This is a system of linear homogeneous equations. It has non-trivial solutions only when its determinant $\Delta$ is equal to zero: $\Delta = 0$. From this condition all the shifts of the eigenfrequencies $(\omega_n - \omega_k^{(0)})$ for degenerate eigenmodes corresponding to a given $n$ are found. After this, all parameters $B_{ns}^{(0)}$ for degenerate eigenmodes can be found from equations (45) with allowance for (39).

For those $k$ for which $\omega_n^{(0)} \neq \omega_k^{(0)}$ (non-degenerate modes), equations (43) allow immediately finding the parameters $B_{nk}^{(1)}$, and hence the parameters $B_{nk}$:

$$B_{nk} = B_{nk}^{(1)} = \frac{\sum_s Q_{ks} B_{ns}^{(0)}}{\hbar(\omega_n^{(0)} - \omega_k^{(0)})} \qquad (46)$$

In order that condition (40) be fulfilled identically (in the linear approximation), it is necessary to take $B_{nk}^{(1)} = 0$ for those $k$ for which $\omega_n^{(0)} = \omega_k^{(0)}$ (degenerate modes).

Thus, in the linear approximation (with respect to perturbations $Q_{ks}$) all the coefficients $B_{nk}$ are uniquely determined and the orthonormal system of functions (21), (26) is constructed.

It follows that the addition of the term $\hat{V}\Psi$ into the Dirac equation removes the last degeneracy, and can describe the Lamb shift.

Because the goal of this paper is not to predict the Lamb shift itself, but to describe the Lamb-Retherford experiments, we will subsequently assume that all eigenfrequencies of the hydrogen atom $\omega_n$, determined from equation (45), are known and different.

## 4. Three-level atom in an oscillating classical electromagnetic field

Let the hydrogen atom be in an external alternating electric field $\mathbf{E}(t) = \mathbf{E}_0 \cos \omega t$, where $\mathbf{E}_0$ and $\omega$ are constants.

Under the action of this field, a nonstationary electric charge distribution of the electron wave arises in the hydrogen atom [9-11], which will have a time-varying electric dipole moment

$$\mathbf{d}(t) = \int \mathbf{r}\rho(t, \mathbf{r}) d\mathbf{r} = -e \int \mathbf{r}|\psi|^2 d\mathbf{r} \qquad (47)$$

According to classical electrodynamics [16], the time-varying electric dipole moment of an electron wave creates around the atom an electric field with strength

$$\mathbf{E}_r = \frac{2}{3c^3} \ddot{\mathbf{d}} \qquad (48)$$

As a result, the atom will be in an alternating electric field, which is a superposition of the external field $\mathbf{E}(t)$ and its self-radiation field (48). Using a gauge transformation, the vector potential of this field can be turned to zero ($\mathbf{A} = 0$), while its scalar potential is as follows [9-11]

$$\Phi = -\mathbf{r}\mathbf{E}_0 \cos \omega t - \frac{2}{3c^3} \mathbf{r}\ddot{\mathbf{d}} \qquad (49)$$

As a result, equation (19) takes the form

$$i\hbar \frac{\partial \psi}{\partial t} = \left( -\frac{1}{2m_e}\Delta - \frac{e^2}{r} + e\mathbf{r}\mathbf{E}_0 \cos \omega t + \frac{2e}{3c^3}\mathbf{r}\ddot{\mathbf{d}} + \hat{U} + \hat{W} \right)\psi \qquad (50)$$

or taking into account equation (47),

$$i\hbar \frac{\partial \psi}{\partial t} = \left( -\frac{1}{2m_e}\Delta - \frac{e^2}{r} + e\mathbf{r}\mathbf{E}_0 \cos \omega t - \frac{2e^2}{3c^3}\mathbf{r}\frac{\partial^3}{\partial t^3}\int \mathbf{r}|\psi|^2 d\mathbf{r} + \hat{U} + \hat{W} \right)\psi \qquad (51)$$

Equation (50), (51) takes into account the corrections associated with the self-radiation field (48). Without taking into account the terms describing the fine structure of the spectral lines of the hydrogen atom and the Lamb shift, this equation was first obtained in [8].

A distinctive feature of equation (51) is that it describes not an abstract field of probabilities, but the deterministic evolution of a single atom in time. This is due to the fact that the field $\psi$ is considered actual, and the function $|\psi|^2$ describes the electric charge distribution of an electron wave in a specific hydrogen atom.

As shown in [9-11], this equation allows describing completely and consistently (without any quantization, in the framework of classical field theory) many quantum processes, including spontaneous emission of the hydrogen atom, a change in its structure (what is traditionally called the quantum transitions) in the process of spontaneous emission, the light-atom interaction, thermal radiation, including the Planck formula and the so-called Einstein $A$-coefficient, etc. In particular, this equation makes it possible to obtain in the framework of classical field theory (i.e. considering only classical fields continuous in space and time) many well-known relations that are traditionally derived in the framework of quantum electrodynamics using quantization of both the atom itself and the radiation field.

It is further shown that equation (51), which takes into account the inverse action of the self-radiation on the electron wave, allows completely describing the Lamb-Retherford experiment in the framework of classical field theory without resorting to quantization of neither the atom nor the radiation field.

The solution of equation (51) will still be sought in the form (20), but with coefficients $c_n$ depending on the time:

$$\psi(t, \mathbf{r}) = \sum_n c_n(t) u_n(\mathbf{r}) \exp(-i\omega_n t) \qquad (52)$$

where the functions $u_n(\mathbf{r})$ are defined by the relation (21) with two-component parameters that satisfy the normalization condition (25), and are solutions of equations (32); thus, the functions $u_n(\mathbf{r})$ form an orthonormal system according to equation (26); the frequencies $\omega_n$ are eigenvalues of the system of equations (30), and the parameters $c_n(t)$ satisfy the normalization condition (27), which expresses only the law of conservation of the charge of the electron wave, which can only be redistributed between the eigenmodes of the atom [9]: the value $-e|c_n(t)|^2$ are equal to the charge of the electron wave in the mode $n$.

Substituting equation (52) into equation (50), taking into account equation (30), we obtain

$$i\hbar \sum_n \frac{dc_n}{dt} u_n(\mathbf{r}) \exp(-i\omega_n t) = \sum_n c_n(t) \exp(-i\omega_n t) \left( \mathbf{E}_0 \cos \omega t + \frac{2}{3c^3} \dddot{\mathbf{d}} \right) e\mathbf{r} u_n(\mathbf{r}) \qquad (53)$$

Multiplying equation (53) by $u_n^*(\mathbf{r}) = (u_{n.1}^* \quad u_{n.2}^*)$, and integrating over the entire space with allowance for (26), we obtain

$$i\hbar \frac{dc_n}{dt} = -\sum_k c_k \mathbf{d}_{nk} \left( \mathbf{E}_0 \cos \omega t + \frac{2}{3c^3} \dddot{\mathbf{d}} \right) \exp(i\omega_{nk} t) \qquad (54)$$

where

$$\omega_{nk} = \omega_n - \omega_k \qquad (55)$$

$$\mathbf{d}_{nk} = \mathbf{d}_{kn}^* = -e \int \mathbf{r} u_n^*(\mathbf{r}) u_k(\mathbf{r}) d\mathbf{r} \tag{56}$$

Taking into account equations (47), (52), (55) and (56), one can write

$$\mathbf{d} = \sum_n \sum_k c_k c_n^* \mathbf{d}_{nk} \exp(i\omega_{nk} t) \tag{57}$$

As shown in [10], the parameters $c_n$ vary with time much more slowly than the oscillating factor $\exp(i\omega_{nk} t)$. This corresponds to the condition

$$|\dot{c}_n| \ll |\omega_{nk}| \tag{58}$$

at least for one pair of excited modes $(n, k)$.

Taking equations (57) and (58) into account, we can approximately write

$$\dddot{\mathbf{d}} = -i \sum_n \sum_k \omega_{nk}^3 c_k c_n^* \mathbf{d}_{nk} \exp(i\omega_{nk} t) \tag{59}$$

For the two-level atom, equations (54) and (59) were analyzed in detail in [9-11].

Let us consider the three-level atom, i.e. atom, in which only three eigenmodes (which we conventionally denote by 1, 2 and 3) are simultaneously excited (do not confuse these numbers with the principal quantum numbers).

For definiteness, we take

$$\omega_1 < \omega_2 < \omega_3 \tag{60}$$

where the mode 1 is always considered as the ground mode $1s_{1/2}$, while the modes 2 and 3 are degenerate from the point of view of the Schrödinger equation, since according to the Schrödinger equation $\omega_2^{(0)} = \omega_3^{(0)}$. However, taking into account high-order effects (fine structure and Lamb shift), the eigenfrequencies $\omega_2$ and $\omega_3$ differ from each other, although much less than from the frequency $\omega_1$, i.e.

$$|\omega_{32}| \ll \omega_{31} \approx \omega_{21} \tag{61}$$

The goal of this paper is to calculate the Lamb-Retherford experiment itself, assuming that there is a small frequency shift $\omega_{32}$ between the "degenerate" eigenmodes. Therefore, in what follows we assume that all frequencies $\omega_{nk}$ are given.

In the case under consideration

$$\dddot{\mathbf{d}} = i\omega_{21}^3 \{c_2 c_1^* \mathbf{d}_{21}^* \exp(-i\omega_{21} t) - c_1 c_2^* \mathbf{d}_{21} \exp(i\omega_{21} t)\} + i\omega_{31}^3 \{c_3 c_1^* \mathbf{d}_{31}^* \exp(-i\omega_{31} t) - c_1 c_3^* \mathbf{d}_{31} \exp(i\omega_{31} t)\} + i\omega_{32}^3 \{c_3 c_2^* \mathbf{d}_{32}^* \exp(-i\omega_{32} t) - c_2 c_3^* \mathbf{d}_{32} \exp(i\omega_{32} t)\} \tag{62}$$

and equations (54) take the form

$$i\hbar \frac{dc_1}{dt} = -c_1 (\mathbf{d}_{11} \mathbf{E}_0) \cos \omega t - c_2 (\mathbf{d}_{12} \mathbf{E}_0) \exp(-i\omega_{21} t) \cos \omega t - c_3 (\mathbf{d}_{13} \mathbf{E}_0) \exp(-i\omega_{31} t) \cos \omega t - \frac{2}{3c^3} \left( c_1 (\mathbf{d}_{11} \dddot{\mathbf{d}}) + c_2 (\mathbf{d}_{12} \dddot{\mathbf{d}}) \exp(-i\omega_{21} t) + c_3 (\mathbf{d}_{13} \dddot{\mathbf{d}}) \exp(-i\omega_{31} t) \right) \tag{63}$$

$$i\hbar\frac{dc_2}{dt} =$$
$$-c_1(\mathbf{d}_{21}\mathbf{E}_0)\exp(i\omega_{21}t)\cos\omega t - c_2(\mathbf{d}_{22}\mathbf{E}_0)\cos\omega t - c_3(\mathbf{d}_{23}\mathbf{E}_0)\exp(-i\omega_{32}t)\cos\omega t -$$
$$\frac{2}{3c^3}\left(c_1(\mathbf{d}_{21}\dddot{\mathbf{d}})\exp(i\omega_{21}t) + c_2(\mathbf{d}_{22}\dddot{\mathbf{d}}) + c_3(\mathbf{d}_{23}\dddot{\mathbf{d}})\exp(-i\omega_{32}t)\right) \quad (64)$$

$$i\hbar\frac{dc_3}{dt} = -c_1(\mathbf{d}_{31}\mathbf{E}_0)\exp(i\omega_{31}t)\cos\omega t - c_2(\mathbf{d}_{32}\mathbf{E}_0)\exp(i\omega_{32}t)\cos\omega t + c_3(\mathbf{d}_{33}\mathbf{E}_0)\cos\omega t -$$
$$\frac{2}{3c^3}\left(c_1(\mathbf{d}_{31}\dddot{\mathbf{d}})\exp(i\omega_{31}t) + c_2(\mathbf{d}_{32}\dddot{\mathbf{d}})\exp(i\omega_{32}t) + c_3(\mathbf{d}_{33}\dddot{\mathbf{d}})\right) \quad (65)$$

Substituting equation (62) into equations (63) - (65), after simple transformations (see Appendix), we obtain

$$i\hbar\frac{dc_1}{dt} = -c_1(\mathbf{d}_{11}\mathbf{E}_0)\cos\omega t + \frac{2}{3c^3}c_2 i\omega_{21}^3 c_1 c_2^*|\mathbf{d}_{21}|^2 + \frac{2}{3c^3}c_2 i\omega_{21}^3 c_1 c_3^*(\mathbf{d}_{21}^*\mathbf{d}_{31})\exp(i\omega_{32}t) +$$
$$\frac{2}{3c^3}c_3 i\omega_{21}^3 c_1 c_2^*(\mathbf{d}_{31}^*\mathbf{d}_{21})\exp(-i\omega_{32}t) + \frac{2}{3c^3}c_3 i\omega_{21}^3 c_1 c_3^*|\mathbf{d}_{31}|^2 \quad (66)$$

$$i\hbar\frac{dc_2}{dt} = -c_2(\mathbf{d}_{22}\mathbf{E}_0)\cos\omega t - c_3(\mathbf{d}_{32}^*\mathbf{E}_0)\exp(-i\omega_{32}t)\cos\omega t - \frac{2}{3c^3}c_1 i\omega_{21}^3 c_2 c_1^*|\mathbf{d}_{21}|^2 -$$
$$\frac{2}{3c^3}c_1 i\omega_{21}^3 c_3 c_1^*(\mathbf{d}_{21}\mathbf{d}_{31}^*)\exp(-i\omega_{32}t) \quad (67)$$

$$i\hbar\frac{dc_3}{dt} =$$
$$-c_2(\mathbf{d}_{32}\mathbf{E}_0)\exp(i\omega_{32}t)\cos\omega t + c_3(\mathbf{d}_{33}\mathbf{E}_0)\cos\omega t - \frac{2}{3c^3}c_1 i\omega_{21}^3 c_2 c_1^*(\mathbf{d}_{31}\mathbf{d}_{21}^*)\exp(i\omega_{32}t) -$$
$$\frac{2}{3c^3}c_1 i\omega_{21}^3 c_3 c_1^*|\mathbf{d}_{31}|^2 \quad (68)$$

Let us consider separately the following cases.

(i) mode 1 is $1s_{1/2}$; mode 2 is $2s_{1/2}$; mode 3 is $2p_{3/2}$;

$$\mathbf{d}_{21} = \mathbf{d}_{21}^* = 0 \quad (69)$$

In this case, because of condition (69), spontaneous emission at the frequency $\omega_{21}$ corresponding to the mixed state 1-2 (i.e., with simultaneous excitation of two eigenmodes 1 and 2) is absent ("forbidden transition") in full accordance with classical electrodynamics [9], while the "transition" $2 \to 3$ (i.e. $2s_{1/2} \to 2p_{3/2}$) induced by microwave radiation, corresponds to the fine structure of the hydrogen atom spectrum.

(ii) mode 1 is $1s_{1/2}$; mode 2 is $2p_{1/2}$; mode 3 is $2s_{1/2}$;

$$\mathbf{d}_{31} = \mathbf{d}_{31}^* = 0 \quad (70)$$

In this case, according to classical electrodynamics, in view of condition (70), spontaneous emission at a frequency $\omega_{31}$ corresponding to mixed state 1-3 (i.e., with simultaneous excitation of two eigenmodes 1 and 3) is absent ("forbidden transition") [9], while the "transition" $3 \to 2$ (i.e., $2s_{1/2} \to 2p_{1/2}$) induced by the microwave radiation, corresponds to the Lamb shift. For the case (69), the equations (66) - (68) take the form

$$\frac{dc_1}{dt} = ic_1 b_{11}\cos\omega t + \gamma_{31}c_1|c_3|^2 \quad (71)$$

$$\frac{dc_2}{dt} = ic_2 b_{22} \cos \omega t + ic_3 b_{32}^* \exp(-i\omega_{32} t) \cos \omega t \tag{72}$$

$$\frac{dc_3}{dt} = ic_2 b_{32} \exp(i\omega_{32} t) \cos \omega t - ic_3 b_{33} \cos \omega t - \gamma_{31} c_3 |c_1|^2 \tag{73}$$

where

$$\gamma_{nk} = \frac{2\omega_{nk}^3}{3\hbar c^3} |\mathbf{d}_{nk}|^2 \tag{74}$$

$$b_{nk} = b_{kn}^* = \frac{1}{\hbar}(\mathbf{d}_{nk}\mathbf{E}_0) \tag{75}$$

We introduce the notation

$$\rho_{nn} = |c_n|^2, \rho_{kk} = |c_k|^2, \rho_{nk} = c_n c_k^*, \rho_{kn} = c_k c_n^* \tag{76}$$

Obviously,

$$\rho_{nk} = \rho_{kn}^* \tag{77}$$

According to equation (27)

$$\rho_{11} + \rho_{22} + \rho_{33} = 1 \tag{78}$$

Then from equations (71) - (73), we obtain

$$\frac{d\rho_{11}}{dt} = 2\gamma_{31}\rho_{11}\rho_{33} \tag{79}$$

$$\frac{d\rho_{22}}{dt} = i\rho_{32} b_{32}^* \exp(-i\omega_{32} t) \cos \omega t - i\rho_{32}^* b_{32} \exp(i\omega_{32} t) \cos \omega t \tag{80}$$

$$\frac{d\rho_{33}}{dt} = i\rho_{32}^* b_{32} \exp(i\omega_{32} t) \cos \omega t - i\rho_{32} b_{32}^* \exp(-i\omega_{32} t) \cos \omega t - 2\gamma_{31}\rho_{11}\rho_{33} \tag{81}$$

$$\frac{d\rho_{32}}{dt} = i(\rho_{22} - \rho_{33})b_{32} \exp(i\omega_{32} t) \cos \omega t - i\rho_{32}(b_{33} + b_{22}) \cos \omega t - \gamma_{31}\rho_{11}\rho_{32} \tag{82}$$

The system of equations (71) - (73) or (79) - (82) completely describes the Lamb-Retherford experiments with respect to the fine structure of the spectral lines of the hydrogen atom.

Taking into account the relation (78), which, as is easy to verify, is a consequence of equations (79) - (81), we can consider a system of only three equations, for example, (79), (81) and (82) or (79), (80 ) and (82).

Further we consider the solution of equations (79) - (82) corresponding to the case (69). We note that for the case (70), equations (66) - (68) can also be rewritten in a form similar to equations (71) - (73) and (79) - (82). However, all results for the case (70) can be obtained formally, using the results obtained for the case (69). To do this, it suffices to interchange the indices 2 and 3 in the solutions of equations (71) - (73) and (79) - (82).

## 5. Calculation of the Lamb-Retherford experiment

We consider an approximate solution of equations (79) - (82), which corresponds to the conditions of the Lamb-Retherford experiments.

In the Lamb-Retherford experiments, all the eigenmodes of the hydrogen atom, except for the ground mode, were excited weakly, which corresponds to the condition

$$\rho_{22}, \rho_{33}, \rho_{32} \ll \rho_{11} \approx 1 \tag{83}$$

In this case the formal solution of equation (82) has the form

$$\rho_{32} = ib_{32} \exp\left(-i\frac{(b_{33}+b_{22})}{\omega}\sin\omega t - \gamma_{31}t\right) \int_0^t (\rho_{22} - \rho_{33}) \cos\omega t' \exp\left(i\frac{(b_{33}+b_{22})}{\omega}\sin\omega t' + \gamma_{31}t' + i\omega_{32}t'\right) dt' + c_0 \exp\left(-i\frac{(b_{33}+b_{22})}{\omega}\sin\omega t - \gamma_{31}t\right) \tag{84}$$

where $c_0$ is an arbitrary constant.

We assume that at the initial instant of time the mode 3 was not excited (even if it was excited, it decays to the ground state in a time of the order of $10^{-9}$ s), i.e., $\rho_{32}(0) = 0$. In this case we obtain $c_0 = 0$ and

$$\rho_{32} = ib_{32} \exp\left(-i\frac{(b_{33}+b_{22})}{\omega}\sin\omega t - \gamma_{31}t\right) \int_0^t (\rho_{22} - \rho_{33}) \cos\omega t' \exp\left(i\frac{(b_{33}+b_{22})}{\omega}\sin\omega t' + \gamma_{31}t' + i\omega_{32}t'\right) dt' \tag{85}$$

We note that in Lamb-Retherford experiments, to observe the effects studied, the oscillation frequency of the external electromagnetic field $\omega$ should have been close to the "transition frequency" $\omega_{32}$. Therefore, in the sequel, we will only be interested in those frequencies $\omega$ that satisfy condition

$$|\omega_{32} - \omega| \ll \omega_{32} \tag{86}$$

We consider weak external actions satisfying condition

$$|b_{33} + b_{22}|/\omega \ll 1 \tag{87}$$

In this case, the function $\left|\frac{(b_{33}+b_{22})}{\omega}\sin\omega t'\right| \ll 1$ at all instants of time, while the parameter $\omega_{32}t$ can significantly exceed unity. This allows neglecting the function $\frac{(b_{33}+b_{22})}{\omega}\sin\omega t'$ in the exponents in the expression (85). In addition, we assume that the function $(\rho_{22} - \rho_{33})$ varies slowly in comparison with the fast oscillating function $\exp(i\omega_{32}t)$. As a result, expression (85) can be rewritten in the form

$$\rho_{32} = i(\rho_{22} - \rho_{33})b_{32}\exp(-\gamma_{31}t) \int_0^t \cos\omega t' \exp(\gamma_{31}t' + i\omega_{32}t') \, dt' \tag{88}$$

After a simple integration, one obtains

$$\rho_{32} = i\frac{1}{2}(\rho_{22} - \rho_{33})b_{32}\exp(-\gamma_{31}t)\left[\frac{\exp(\gamma_{31}t+i\omega_{32}t+i\omega t)-1}{(\gamma_{31}+i\omega_{32}+i\omega)} + \frac{\exp(\gamma_{31}t+i\omega_{32}t-i\omega t)-1}{(\gamma_{31}+i\omega_{32}-i\omega)}\right] \tag{89}$$

Substituting equation (89) into equation (80), we obtain

$$\frac{d\rho_{22}}{dt} = -|b_{32}|^2(\rho_{22} - \rho_{33})\left\{\frac{1}{4}\left\{\frac{\exp(2i\omega t)-\exp(-\gamma_{31}t-i\omega_{32}t+i\omega t)}{(\gamma_{31}+i\omega_{32}+i\omega)} + \frac{1-\exp(-\gamma_{31}t-i\omega_{32}t+i\omega t)}{(\gamma_{31}+i\omega_{32}-i\omega)} + \right.\right.$$

$$\frac{1-\exp(-\gamma_{31}t+i\omega_{32}t+i\omega t)}{(\gamma_{31}-i\omega_{32}-i\omega)} + \frac{\exp(2i\omega t)-\exp(-\gamma_{31}t+i\omega_{32}t+i\omega t)}{(\gamma_{31}-i\omega_{32}+i\omega)}\right\} + \frac{1}{4}\left\{\frac{1-\exp(-\gamma_{31}t-i\omega_{32}t-i\omega t)}{(\gamma_{31}+i\omega_{32}+i\omega)} + \right.$$

$$\left.\left.\frac{\exp(-2i\omega t)-\exp(-\gamma_{31}t-i\omega_{32}t-i\omega t)}{(\gamma_{31}+i\omega_{32}-i\omega)} + \frac{\exp(-2i\omega t)-\exp(-\gamma_{31}t+i\omega_{32}t-i\omega t)}{(\gamma_{31}-i\omega_{32}-i\omega)} + \frac{1-\exp(-\gamma_{31}t+i\omega_{32}t-i\omega t)}{(\gamma_{31}-i\omega_{32}+i\omega)}\right\}\right\} \quad (90)$$

Taking into account that the function $\rho_{22}$ is assumed to be slowly varying in comparison with the fast oscillating function $\exp(2i\omega_{32}t)$, equation (90) can be averaged over times of the order of $\omega_{32}^{-1}$. In this case, in view of condition (86), the exponents containing $(\omega_{32} - \omega)$ must be considered to be constant. As a result, one obtains

$$\frac{d\rho_{22}}{dt} = -\frac{1}{2}|b_{32}|^2(\rho_{22} - \rho_{33})\left\{\frac{\gamma_{31}}{\gamma_{31}^2+(\omega_{32}+\omega)^2} + \frac{\gamma_{31}}{\gamma_{31}^2+(\omega_{32}-\omega)^2} - \exp(-\gamma_{31}t)\left[\frac{(\gamma_{31}-i\omega_{32})}{(\gamma_{31}-i\omega_{32})^2+\omega^2}\exp(i\omega_{32}t - i\omega t) + \frac{(\gamma_{31}+i\omega_{32})}{(\gamma_{31}+i\omega_{32})^2+\omega^2}\exp(-i\omega_{32}t + i\omega t)\right]\right\} \quad (91)$$

In addition, we take into account that the mixed excited state of a hydrogen atom with two excited modes, $n = 1$ and $n = 3$, is unstable, and the electric charge of the electron wave rapidly leaves the mode $n = 3$ due to flow into the ground mode $n = 1$. In this case, we can approximately take

$$\rho_{22} - \rho_{33} \approx \rho_{22} \quad (92)$$

As a result, equation (91) takes the form

$$\frac{d\rho_{22}}{dt} = -\frac{1}{2}|b_{32}|^2\rho_{22}\left\{\frac{\gamma_{31}}{\gamma_{31}^2+(\omega_{32}+\omega)^2} + \frac{\gamma_{31}}{\gamma_{31}^2+(\omega_{32}-\omega)^2} - \exp(-\gamma_{31}t)\left[\frac{(\gamma_{31}-i\omega_{32})}{(\gamma_{31}-i\omega_{32})^2+\omega^2}\exp(i\omega_{32}t - i\omega t) - \frac{(\gamma_{31}+i\omega_{32})}{(\gamma_{31}+i\omega_{32})^2+\omega^2}\exp(-i\omega_{32}t + i\omega t)\right]\right\} \quad (93)$$

The solution of equation (93) is as follows

$$\rho_{22} = \rho_{22}(0)\exp\left\{-\frac{1}{2}|b_{32}|^2\left[\frac{\gamma_{31}t}{\gamma_{31}^2+(\omega_{32}+\omega)^2} + \frac{\gamma_{31}t}{\gamma_{31}^2+(\omega_{32}-\omega)^2} - \sqrt{\frac{(\gamma_{31}^2+\omega_{32}^2)}{[(\gamma_{31}^2+\omega_{32}^2)^2+2\omega^2(\gamma_{31}^2-\omega_{32}^2)+\omega^4][\gamma_{31}^2+(\omega_{32}-\omega)^2]}}\exp(-\gamma_{31}t)\sin[i(\omega_{32} - \omega)t]\right]\right\} \quad (94)$$

Calculations show that the function of time in the exponent is close to linear, and with sufficient accuracy one can write

$$\rho_{22} = \rho_{22}(0)\exp\left\{-\frac{|b_{32}|^2}{2\gamma_{31}}\lambda t\right\} \quad (95)$$

where

$$\lambda = \frac{\gamma_{31}^2}{\gamma_{31}^2+(\omega_{32}+\omega)^2} + \frac{\gamma_{31}^2}{\gamma_{31}^2+(\omega_{32}-\omega)^2} \qquad (96)$$

Thus, under the action of an external periodic electric field, an irreversible induced flow of electron wave from the metastable mode $n = 2$ into the unstable rapidly decaying mode $n = 3$ occurs.

Dependences of the nondimensional decrement $\lambda$ on the nondimensional frequency $\omega/\gamma_{31}$ of the external field for different values of the nondimensional "transition frequency" $\omega_{32}/\gamma_{31}$ are shown in Fig. 1.

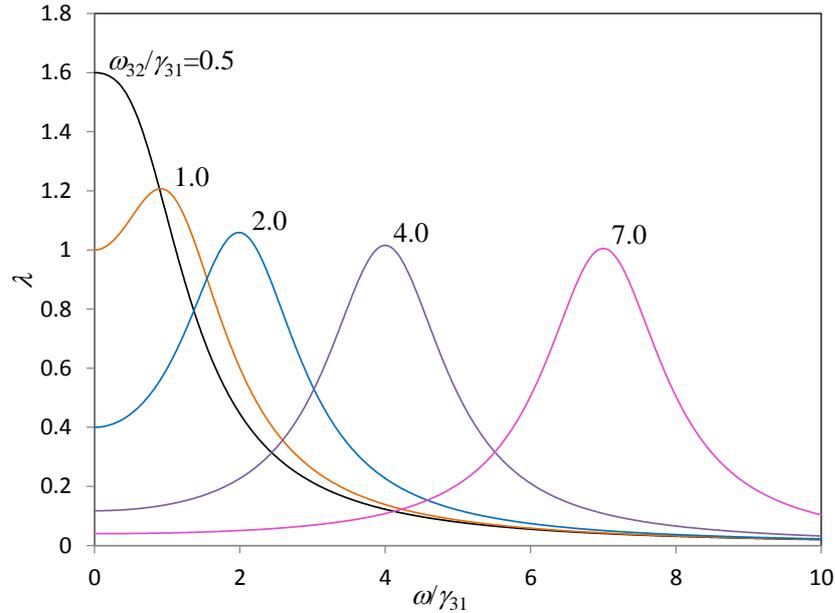

Fig. 1. (Colour online) Dependences of the nondimensional decrement $\lambda$ on the nondimensional frequency $\omega/\gamma_{31}$ of the external field for different values of the nondimensional "transition frequency" $\omega_{32}/\gamma_{31}$.

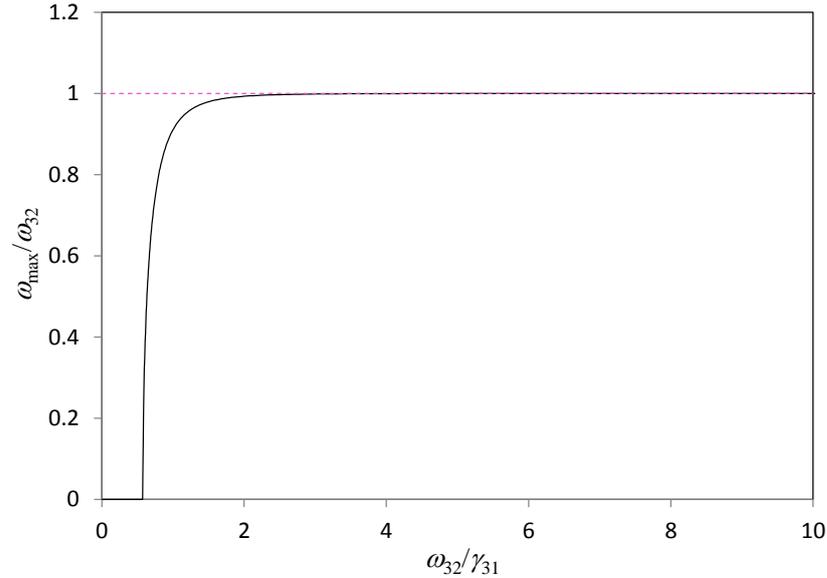

Fig. 2. (Colour online) Dependence of frequency $\omega_{\max}$, at which the maximum of the damping decrement $\lambda$ is reached, on the nondimensional "transition frequency" $\omega_{32}/\gamma_{31}$.

As follows from Fig. 1, when $\omega_{32}/\gamma_{31} > 1$, the dependence of the decrement $\lambda$ on the frequency $\omega$ of the external field has a pronounced maximum. The frequency $\omega_{max}$ at which this maximum is reached is determined by the relation

$$\omega_{max}^2 = -(\gamma_{31}^2 + \omega_{32}^2) \pm \sqrt{(\gamma_{31}^2 + \omega_{32}^2)^2 + 2\omega_{32}^2\gamma_{31}^2 - \gamma_{31}^4 + 3\omega_{32}^4} \qquad (97)$$

The dependence of the nondimensional frequency $\omega_{max}/\omega_{32}$ on the parameter $\omega_{32}/\gamma_{31}$ is shown in Fig. 2.

It is seen that for $\omega_{32}/\gamma_{31} < 0.575$ the maximum of the decrement $\lambda$ is observed at $\omega = 0$, and at $0.57 < \omega_{32}/\gamma_{31} < 1.5$ the frequency $\omega_{max}$ is less than the "transition frequency" $\omega_{32}$ and rapidly decreases to zero with decrease in the parameter $\omega_{32}/\gamma_{31}$. And only at $\omega_{32}/\gamma_{31} > 1.5$ with sufficient accuracy one can consider that $\omega_{max} \approx \omega_{32}$, and the larger the ratio $\omega_{32}/\gamma_{31}$ the more exactly. In particular, for the shift of the eigenfrequencies corresponding to the fine structure of the spectral line $2s_{1/2} - 2p_{3/2}$ of the hydrogen atom, $\omega_{32}/\gamma_{31} \approx 200$, while for the Lamb shift of the $2s_{1/2}$ and $2p_{1/2}$ levels in the hydrogen atom, $\omega_{32}/\gamma_{21} \approx 21$. Thus, in both cases the condition $\omega_{max} = \omega_{32}$ is fulfilled with high accuracy.

This means that when an external periodic field with an appropriate frequency is applied to such an atom, the electric charge of the electron wave flows from the metastable $2s_{1/2}$ mode into the unstable modes $2p_{1/2}$ or $2p_{3/2}$, which quickly and spontaneously relax to the ground mode $1s_{1/2}$; in this case the maximum velocity of the process is observed at $\omega = \omega_{32}$ which allowed to Lamb and Retherford correctly determining the frequency shifts $\omega_{32}$.

It is interesting to note that if it would be $\omega_{32}/\gamma_{31} \leq 0.57$, then the frequency shift $\omega_{32}$ would not be fixed in the Lamb-Retherford experiments, because the maximum rate of stimulated decay of the $2s_{1/2}$ mode would be observed at $\omega = 0$, while at $0.57 < \omega_{32}/\gamma_{31} < 1.5$ the result of the Lamb-Retherford experiments would give a significant error in comparison with the actual value $\omega_{32}$. Thus, the correct result for the Lamb shift was obtained in the Lamb-Retherford experiments only because of the fortunate circumstance that for the Lamb shift of the $2s_{1/2}$ and $2p_{1/2}$ levels in the hydrogen atom there is a large ratio $\omega_{32}/\gamma_{21} \approx 21$.

## 6. Quasi-stationary approximation

Let us consider another method for solving equations (79) - (82). We assume that the parameter $\rho_{33}$ varies slowly (quasi-statically). In this case we take

$$\frac{d\rho_{33}}{dt} \approx 0 \qquad (98)$$

Then the equations (79) - (81) take the form

$$i\rho_{32}^* b_{32} \exp(i\omega_{32}t) \cos\omega t - i\rho_{32} b_{32}^* \exp(-i\omega_{32}t) \cos\omega t = 2\gamma_{31}\rho_{11}\rho_{33} \qquad (99)$$

$$\frac{d\rho_{11}}{dt} = -\frac{d\rho_{22}}{dt} = 2\gamma_{31}\rho_{11}\rho_{33} \tag{100}$$

From equation (100), it follows how much of electric charge of the electron wave flows out of the mode 2, the same amount of electric charge flows into the mode 1. In other words, in the hydrogen atom the internal electric current 2→3 (associated with the flow of electric charge of the electron wave from mode 2 into mode 3) is equal to the internal electric current 3→1 and is equal to $-e2\gamma_{31}\rho_{11}\rho_{33}$.

Let us substitute the approximate solution of equation (82) in the form (89) into equation (99). As a result, after integrating and averaging over fast oscillations with allowance for (86), we obtain

$$2\gamma_{31}\rho_{11}\rho_{33} = \frac{1}{2}\gamma_{31}|b_{32}|^2(\rho_{22} - \rho_{33})\left\{\frac{1}{\gamma_{31}^2 + (\omega_{32} - i\omega)^2} + \frac{1}{\gamma_{31}^2 + (\omega_{32} + i\omega)^2}\right\} \tag{101}$$

or, taking equation (96) into account, one obtains

$$2\gamma_{31}\rho_{11}\rho_{33} = \lambda\frac{|b_{32}|^2}{2\gamma_{31}}(\rho_{22} - \rho_{33}) \tag{102}$$

In the approximation (86) and (92) we obtain

$$\rho_{33} = \lambda\frac{|b_{32}|^2}{4\gamma_{31}^2}\rho_{22} \tag{103}$$

Substituting equation (103) into equation (100) one obtains

$$\frac{d\rho_{22}}{dt} = -\lambda\frac{|b_{32}|^2}{2\gamma_{31}}\rho_{22} \tag{104}$$

The solution of equation (104) has the form (95), and, therefore, all the results and conclusions obtained above remain valid in the quasi-stationary approximation (98). Thus, the approximate solution considered in Section 5 corresponds to the quasi-stationary changing in the mode $n = 3$, which according to (103) instantaneously adapts to the current state of the mode $n = 2$.

We note that the condition (92) means that $\rho_{33} \ll \rho_{22}$; in this case taking into account relation (103), the condition

$$\lambda\frac{|b_{32}|^2}{4\gamma_{31}^2} \ll 1 \tag{105}$$

must be satisfied.

The condition of slow variation of the functions $\rho_{22}$ and $\rho_{33}$ in comparison with the fast oscillating function $\exp(i\omega_{32}t)$ means that the inequality $\left|\frac{d\rho_{22}}{dt}\right| \ll \omega_{32}\rho_{22}$ must be satisfied. Taking equation (104) into account, we obtain the condition for a slow change in the functions $\rho_{22}$ and $\rho_{33}$ in the form

$$\lambda\frac{|b_{32}|^2}{2\gamma_{31}\omega_{32}} \ll 1 \tag{106}$$

Taking into account that for the effects under consideration, $\gamma_{31}/\omega_{32} \ll 1$, we come to the conclusion that the condition (106) is a consequence of condition (105). Differentiating equation (103) with respect to time, we obtain

$$\frac{d\rho_{33}}{dt} = \lambda \frac{|b_{32}|^2}{4\gamma_{31}^2} \frac{d\rho_{22}}{dt} \tag{107}$$

Taking into account condition (105), from equation (107) we obtain

$$\frac{d\rho_{33}}{dt} \ll \frac{d\rho_{22}}{dt} \tag{108}$$

which is a justification of the quasi-stationary approximation (98).

## 7. Concluding remarks

We note that a magnetic field was used in the Lamb-Retherford experiments, which simultaneously performed several functions (see Introduction). In particular, it has allowed increasing the frequency shift due to the anomalous Zeeman effect, and, thereby, increasing the sensitivity of the method. In this paper, the magnetic field was not explicitly taken into account, but it can be taken into account by simply adding the corresponding Zeeman additive to the frequency shift $\omega_{32}$.

Thus, we have shown that Lamb-Retherford experiments can be described in natural way within the framework of classical field theory without any quantization if we take into account the inverse effect of the self-radiation field of an atom which is in a mixed excited state.

This paper extends the class of phenomena described by the theory [8-14].

**Acknowledgments**

Funding was provided by the Tomsk State University competitiveness improvement program.

**Appendix**

Substituting equation (62) into equations (63) - (65), after simple transformations we obtain

$$i\hbar \frac{dc_1}{dt} = -c_1(\mathbf{d}_{11}\mathbf{E}_0)\cos\omega t - c_2(\mathbf{d}_{21}^*\mathbf{E}_0)\exp(-i\omega_{21}t)\cos\omega t - c_3(\mathbf{d}_{31}^*\mathbf{E}_0)\exp(-i\omega_{31}t)\cos\omega t -$$
$$- \frac{2}{3c^3} c_1 i\omega_{21}^3 \{c_2 c_1^*(\mathbf{d}_{11}\mathbf{d}_{21}^*)\exp(-i\omega_{21}t) - c_1 c_2^*(\mathbf{d}_{11}\mathbf{d}_{21})\exp(i\omega_{21}t)\} -$$
$$- \frac{2}{3c^3} c_1 i\omega_{31}^3 \{c_3 c_1^*(\mathbf{d}_{11}\mathbf{d}_{31}^*)\exp(-i\omega_{31}t) - c_1 c_3^*(\mathbf{d}_{11}\mathbf{d}_{31})\exp(i\omega_{31}t)\} -$$
$$- \frac{2}{3c^3} c_1 i\omega_{32}^3 \{c_3 c_2^*(\mathbf{d}_{11}\mathbf{d}_{32}^*)\exp(-i\omega_{32}t) - c_2 c_3^*(\mathbf{d}_{11}\mathbf{d}_{32})\exp(i\omega_{32}t)\} -$$
$$- \frac{2}{3c^3} c_2 i\omega_{21}^3 \{c_2 c_1^*(\mathbf{d}_{21}^*\mathbf{d}_{21}^*)\exp(-2i\omega_{21}t) - c_1 c_2^*(\mathbf{d}_{21}^*\mathbf{d}_{21})\} -$$

$$\frac{2}{3c^3} c_2 i\omega_{31}^3 \{c_3 c_1^*(\mathbf{d}_{21}^* \mathbf{d}_{31}^*) \exp(-i(\omega_{31}+\omega_{21})t) - c_1 c_3^*(\mathbf{d}_{21}^* \mathbf{d}_{31}) \exp(i\omega_{32}t)\} -$$

$$\frac{2}{3c^3} c_2 i\omega_{32}^3 \{c_3 c_2^*(\mathbf{d}_{21}^* \mathbf{d}_{32}^*) \exp(-i\omega_{31}t) - c_2 c_3^*(\mathbf{d}_{21}^* \mathbf{d}_{32}) \exp(i(\omega_{32}-\omega_{21})t)\} -$$

$$\frac{2}{3c^3} c_3 i\omega_{21}^3 \{c_2 c_1^*(\mathbf{d}_{31}^* \mathbf{d}_{21}^*) \exp(-i(\omega_{21}+\omega_{31})t) - c_1 c_2^*(\mathbf{d}_{31}^* \mathbf{d}_{21}) \exp(-i\omega_{32}t)\} -$$

$$\frac{2}{3c^3} c_3 i\omega_{31}^3 \{c_3 c_1^*(\mathbf{d}_{31}^* \mathbf{d}_{31}^*) \exp(-2i\omega_{31}t) - c_1 c_3^*(\mathbf{d}_{31}^* \mathbf{d}_{31})\} -$$

$$\frac{2}{3c^3} c_3 i\omega_{32}^3 \{c_3 c_2^*(\mathbf{d}_{31}^* \mathbf{d}_{32}^*) \exp(-i(\omega_{32}+\omega_{31})t) - c_2 c_3^*(\mathbf{d}_{31}^* \mathbf{d}_{32}) \exp(-i\omega_{21}t)\} \quad (A1)$$

$$i\hbar \frac{dc_2}{dt} =$$

$$-c_1(\mathbf{d}_{21}\mathbf{E}_0) \exp(i\omega_{21}t) \cos\omega t - c_2(\mathbf{d}_{22}\mathbf{E}_0) \cos\omega t - c_3(\mathbf{d}_{32}^*\mathbf{E}_0) \exp(-i\omega_{32}t) \cos\omega t -$$

$$\frac{2}{3c^3} c_1 i\omega_{21}^3 \{c_2 c_1^*(\mathbf{d}_{21}\mathbf{d}_{21}^*) - c_1 c_2^*(\mathbf{d}_{21}\mathbf{d}_{21}) \exp(2i\omega_{21}t)\} -$$

$$\frac{2}{3c^3} c_1 i\omega_{31}^3 \{c_3 c_1^*(\mathbf{d}_{21}\mathbf{d}_{31}^*) \exp(-i\omega_{32}t) - c_1 c_3^*(\mathbf{d}_{21}\mathbf{d}_{31}) \exp(i(\omega_{31}+\omega_{21})t)\} -$$

$$\frac{2}{3c^3} c_1 i\omega_{32}^3 \{c_3 c_2^*(\mathbf{d}_{21}\mathbf{d}_{32}^*) \exp(i(\omega_{21}-\omega_{32})t) - c_2 c_3^*(\mathbf{d}_{21}\mathbf{d}_{32}) \exp(i\omega_{31}t)\} -$$

$$\frac{2}{3c^3} c_2 i\omega_{21}^3 \{c_2 c_1^*(\mathbf{d}_{22}\mathbf{d}_{21}^*) \exp(-i\omega_{21}t) - c_1 c_2^*(\mathbf{d}_{22}\mathbf{d}_{21}) \exp(i\omega_{21}t)\} -$$

$$\frac{2}{3c^3} c_2 i\omega_{31}^3 \{c_3 c_1^*(\mathbf{d}_{22}\mathbf{d}_{31}^*) \exp(-i\omega_{31}t) - c_1 c_3^*(\mathbf{d}_{22}\mathbf{d}_{31}) \exp(i\omega_{31}t)\} -$$

$$\frac{2}{3c^3} c_2 i\omega_{32}^3 \{c_3 c_2^*(\mathbf{d}_{22}\mathbf{d}_{32}^*) \exp(-i\omega_{32}t) - c_2 c_3^*(\mathbf{d}_{22}\mathbf{d}_{32}) \exp(i\omega_{32}t)\} -$$

$$\frac{2}{3c^3} c_3 i\omega_{21}^3 \{c_2 c_1^*(\mathbf{d}_{32}^* \mathbf{d}_{21}^*) \exp(-i(\omega_{21}+\omega_{32})t) - c_1 c_2^*(\mathbf{d}_{32}^* \mathbf{d}_{21}) \exp(i(\omega_{21}-\omega_{32})t)\} -$$

$$\frac{2}{3c^3} c_3 i\omega_{31}^3 \{c_3 c_1^*(\mathbf{d}_{32}^* \mathbf{d}_{31}^*) \exp(-i(\omega_{31}+\omega_{32})t) - c_1 c_3^*(\mathbf{d}_{32}^* \mathbf{d}_{31}) \exp(i\omega_{21}t)\} -$$

$$\frac{2}{3c^3} c_3 i\omega_{32}^3 \{c_3 c_2^*(\mathbf{d}_{32}^* \mathbf{d}_{32}^*) \exp(-2i\omega_{32}t) - c_2 c_3^*(\mathbf{d}_{32}^* \mathbf{d}_{32})\} \quad (A2)$$

$$i\hbar \frac{dc_3}{dt} = -c_1(\mathbf{d}_{31}\mathbf{E}_0) \exp(i\omega_{31}t) \cos\omega t - c_2(\mathbf{d}_{32}\mathbf{E}_0) \exp(i\omega_{32}t) \cos\omega t + c_3(\mathbf{d}_{33}\mathbf{E}_0) \cos\omega t -$$

$$\frac{2}{3c^3} c_1 i\omega_{21}^3 \{c_2 c_1^*(\mathbf{d}_{31}\mathbf{d}_{21}^*) \exp(i\omega_{32}t) - c_1 c_2^*(\mathbf{d}_{31}\mathbf{d}_{21}) \exp(i(\omega_{21}+\omega_{31})t)\} -$$

$$\frac{2}{3c^3} c_1 i\omega_{31}^3 \{c_3 c_1^*(\mathbf{d}_{31}\mathbf{d}_{31}^*) - c_1 c_3^*(\mathbf{d}_{31}\mathbf{d}_{31}) \exp(2i\omega_{31}t)\} -$$

$$\frac{2}{3c^3} c_1 i\omega_{32}^3 \{c_3 c_2^*(\mathbf{d}_{31}\mathbf{d}_{32}^*) \exp(i\omega_{21}t) - c_2 c_3^*(\mathbf{d}_{31}\mathbf{d}_{32}) \exp(i(\omega_{32}+\omega_{31})t)\} -$$

$$\frac{2}{3c^3} c_2 i\omega_{21}^3 \{c_2 c_1^*(\mathbf{d}_{32}\mathbf{d}_{21}^*) \exp(i(\omega_{32}-\omega_{21})t) - c_1 c_2^*(\mathbf{d}_{32}\mathbf{d}_{21}) \exp(i\omega_{31}t)\} -$$

$$\frac{2}{3c^3} c_2 i\omega_{31}^3 \{c_3 c_1^*(\mathbf{d}_{32}\mathbf{d}_{31}^*) \exp(-i\omega_{21}t) - c_1 c_3^*(\mathbf{d}_{32}\mathbf{d}_{31}) \exp(i(\omega_{31}+\omega_{32})t)\} -$$

$$\frac{2}{3c^3} c_2 i\omega_{32}^3 \{c_3 c_2^*(\mathbf{d}_{32}\mathbf{d}_{32}^*) - c_2 c_3^*(\mathbf{d}_{32}\mathbf{d}_{32}) \exp(2i\omega_{32}t)\} -$$

$$\frac{2}{3c^3} c_3 i\omega_{21}^3 \{c_2 c_1^*(\mathbf{d}_{33}\mathbf{d}_{21}^*) \exp(-i\omega_{21}t) - c_1 c_2^*(\mathbf{d}_{33}\mathbf{d}_{21}) \exp(i\omega_{21}t)\} -$$

$$\frac{2}{3c^3} c_3 i\omega_{31}^3 \{c_3 c_1^*(\mathbf{d}_{33}\mathbf{d}_{31}^*) \exp(-i\omega_{31}t) - c_1 c_3^*(\mathbf{d}_{33}\mathbf{d}_{31}) \exp(i\omega_{31}t)\} -$$

$$\frac{2}{3c^3} c_3 i\omega_{32}^3 \{c_3 c_2^*(\mathbf{d}_{33}\mathbf{d}_{32}^*) \exp(-i\omega_{32}t) - c_2 c_3^*(\mathbf{d}_{33}\mathbf{d}_{32}) \exp(i\omega_{32}t)\} \quad (A3)$$

Taking equations (61) and (87) into account, we assume that

$$\omega \sim \omega_{32} \ll \omega_{21} \sim \omega_{31} \tag{A4}$$

This corresponds to the conditions of the Lamb-Retherford experiment.

Then, averaging equations (A1) - (A3) in terms of fast oscillations, we obtain

$$i\hbar \frac{dc_1}{dt} = -c_1(\mathbf{d}_{11}\mathbf{E}_0)\cos\omega t - \frac{2}{3c^3} c_1 i\omega_{32}^3 \{c_3 c_2^*(\mathbf{d}_{11}\mathbf{d}_{32}^*)\exp(-i\omega_{32}t) - c_2 c_3^*(\mathbf{d}_{11}\mathbf{d}_{32})\exp(i\omega_{32}t)\} +$$
$$\frac{2}{3c^3} c_2 i\omega_{21}^3 c_1 c_2^* |\mathbf{d}_{21}|^2 + \frac{2}{3c^3} c_2 i\omega_{31}^3 c_1 c_3^*(\mathbf{d}_{21}^*\mathbf{d}_{31})\exp(i\omega_{32}t) +$$
$$\frac{2}{3c^3} c_3 i\omega_{21}^3 c_1 c_2^*(\mathbf{d}_{31}^*\mathbf{d}_{21})\exp(-i\omega_{32}t) + \frac{2}{3c^3} c_3 i\omega_{31}^3 c_1 c_3^* |\mathbf{d}_{31}|^2 \tag{A5}$$

$$i\hbar \frac{dc_2}{dt} = -c_2(\mathbf{d}_{22}\mathbf{E}_0)\cos\omega t - c_3(\mathbf{d}_{32}^*\mathbf{E}_0)\exp(-i\omega_{32}t)\cos\omega t - \frac{2}{3c^3} c_1 i\omega_{21}^3 c_2 c_1^* |\mathbf{d}_{21}|^2 -$$
$$\frac{2}{3c^3} c_1 i\omega_{31}^3 c_3 c_1^*(\mathbf{d}_{21}\mathbf{d}_{31}^*)\exp(-i\omega_{32}t) - \frac{2}{3c^3} c_2 i\omega_{32}^3 \{c_3 c_2^*(\mathbf{d}_{22}\mathbf{d}_{32}^*)\exp(-i\omega_{32}t) -$$
$$c_2 c_3^*(\mathbf{d}_{22}\mathbf{d}_{32})\exp(i\omega_{32}t)\} + \frac{2}{3c^3} c_3 i\omega_{32}^3 c_2 c_3^* |\mathbf{d}_{32}|^2 \tag{A6}$$

$$i\hbar \frac{dc_3}{dt} = -c_2(\mathbf{d}_{32}\mathbf{E}_0)\exp(i\omega_{32}t)\cos\omega t + c_3(\mathbf{d}_{33}\mathbf{E}_0)\cos\omega t - \frac{2}{3c^3} c_1 i\omega_{21}^3 c_2 c_1^*(\mathbf{d}_{31}\mathbf{d}_{21}^*)\exp(i\omega_{32}t) -$$
$$\frac{2}{3c^3} c_1 i\omega_{31}^3 c_3 c_1^* |\mathbf{d}_{31}|^2 - \frac{2}{3c^3} c_2 i\omega_{32}^3 c_3 c_2^* |\mathbf{d}_{32}|^2 - \frac{2}{3c^3} c_3 i\omega_{32}^3 \{c_3 c_2^*(\mathbf{d}_{33}\mathbf{d}_{32}^*)\exp(-i\omega_{32}t) -$$
$$c_2 c_3^*(\mathbf{d}_{33}\mathbf{d}_{32})\exp(i\omega_{32}t)\} \tag{A7}$$

Taking into account condition (A4), the terms with the factor $\omega_{32}^3$ will be much less than the terms with the factors $\omega_{31}^3$ and $\omega_{21}^3$. In addition, we can approximately accept

$$\omega_{31}^3 \approx \omega_{21}^3 \tag{A8}$$

As a result, we obtain equations (66) - (68).